# Healing of the edge magnetic island in the island divertor configuration on J-TEXT

**Zhangrong Hou**[1], **Song Zhou**[1,*], **Nengchao Wang**[1], **Yonghua Ding**[1], **Zhonghe Jiang**[1], **Yunfeng Liang**[1,2,3], **Zhengkang Ren**[1], **Feiyue Mao**[1], **Qinghu Yang**[1], **Jiaming Wang**[1], **Xin Xu**[1], **Yutong Yang**[1,2], **Jiankun Hua**[1,2], **Zijian Xuan**[1], **Chuanxu Zhao**[1], **Yangbo Li**[1], **Lei Yu**[1], **Donghui Xia**[1], **Zhipeng Chen**[1], **Zhoujun Yang**[1] and the J-TEXT team[4]

[1] State Key Laboratory of Advanced Electromagnetic Technology, International Joint Research Laboratory of Magnetic Confinement Fusion and Plasma Physics, School of Electrical and Electronic Engineering, Huazhong University of Science and Technology, Wuhan 430074, China
[2] Forschungszentrum Jülich GmbH, Institut für Energie- und Klimaforschung—Plasmaphysik,52425 Jülich, Germany
[3] Institute of Plasma Physics, Chinese Academy of Sciences, Hefei 230031, China

E-mail: szhou@hust.edu.cn

**Abstract**
The phenomena of island healing and configuration transition induced by high-power electron cyclotron resonance heating (ECRH) have been investigated in the island divertor configuration on the J-TEXT tokamak. Experimental results reveal that the size of the edge open magnetic island with mode number $m/n$ = 3/1 decreases substantially under specific ECRH conditions. This process, referred to as island healing, occurs when ECRH with a power of 500~600 kW is deposited in the plasma core or when 250 kW of ECRH is deposited at $r_{EC}$ = 0.5 $a$, where $a$ is the minor radius. The reduction of the island width makes the island divertor ineffective and transition into the limiter configuration. A model incorporating the influence of ECRH on the scrape-off layer (SOL) thermoelectric current is proposed to explain the observed changes in the edge magnetic topology of the island divertor configuration. These findings suggest that ECRH should be deposited at the plasma core with carefully controlled power to ensure the stable and compatible operation of ECRH and the island divertor configuration in tokamaks. The results can provide insights into achieving robust operation of an island divertor in tokamaks.

Keywords: island divertor configuration, tokamak, ECRH, island healing (Some figures may appear in color only in the online journal)

---

## 1. Introduction

Mitigation of the divertor heat load has been regarded as one key issue to realize a high-performance thermal-nuclear fusion reactor. The island divertor configuration was initially proposed for tokamaks as an alternative 3D advanced configuration for tolerable divertor heat exhaust [1]. Over time, it has become a primary approach for enhancing heat exhaust capabilities in stellarators, as demonstrated in devices such as W7-AS [2], LHD [3], HSX [4] and W7-X [5]. First operation with an island divertor in W7-X [5] has brought results about the heat flux reduction, which was at least a factor of 10, and was characterized by a significant drop in target electron temperature. First application of the island divertor configuration in the J-TEXT tokamak achieved a peak heat load reduction of approximately 50% [6]. The application of the 3D magnetic topology of the island divertor configuration can result in a significant change of the scrape-off layer (SOL) transport, causing the improvement of heat exhaust in tokamak operations.

To achieve high-temperature plasmas in the advanced island divertor configuration, auxiliary heating is essential. The reduction of the island size with increasing $\beta$ has been found in the process of the equilibrium calculations for Helias [7]. The magnetic islands, in some cases, show a property of 'self-healing', i.e., they tend to shrink as $\beta$ increases. In the previous studies [8–10] in the LHD and TJ-II stellarators, self-healing of the magnetic island has been observed. Hegna [11] proposed a model to explain the island healing phenomenon in stellarators, attributing it to plasma flows. In this model, the neoclassical transport can provide plasma flows of sufficient magnitude to heal vacuum magnetic islands. Due to the smaller value of neoclassical transport in tokamaks, it is less likely to heal islands in a tokamak than in a stellarator, based on his theory. As such, this model is not suitable for tokamaks.

However, the island divertor in tokamaks may encounter challenges related to the island size. Specifically, the effectiveness of the island divertor configuration could deteriorate if the island size reduces as $\beta$ increases. This raises the need to assess the island divertor's performance under high-power plasma operation. Electron cyclotron resonance heating (ECRH), one of the most efficient auxiliary heating methods, has been implemented on J-TEXT. On this device, significant efforts have been made to investigating the compatibility between the island divertor configuration and high-power ECRH system.

In the present island divertor experiments of the J-TEXT tokamak, it has been found, for the first time, that the width of the edge magnetic island decreases, leading to a transition from the island divertor configuration to the limiter configuration, with different power of on-axis and off-axis ECRH. The remainder of this paper is organized as follows: The edge magnetic topology of the 3/1 island divertor configuration and the relevant diagnostics in J-TEXT are described in section 2; section 3 presents the experimental results of the operation of the ECRH and the island divertor configuration; sections 4 discusses the possible physical mechanism of ECRH's effect on the edge island topology and numerically reproduces the phenomena; and section 5 gives the conclusion.

## 2  Experimental setup

Experiments are performed on J-TEXT, a circular, medium-sized iron core tokamak which features main parameters of (major radius $R_0$ = 1.05 m, minor radius $a$ = 0.21~0.29 m, toroidal magnetic field $B_t$ ~ 2 T), and can produce $I_p$ ~ 200 kA for a discharge time of 400 ms [12]. The J-TEXT tokamak is equipped with four graphite rail limiters. The main limiter is installed at the high field side (HFS) of port 6 with a toroidal angel from 125° to 145° and a distance of $a$ = 0.225 m from the center of the vacuum chamber, which decides the minor radius of plasma. The Ohmic hydrogen discharge parameters for the experiment are: the plasma current $I_p$ = 120-140 kA, the toroidal magnetic field $B_t$ = 1.875 T, the electron density of the core plasma $n_e$ = (2.5~3.0)×10$^{19}$ m$^{-3}$ and the edge safety factor $q_a$ = 3.1~3.6.

*2.1. Edge magnetic topology of the island divertor configuration*

J-TEXT tokamak is equipped with a set of resonant magnetic perturbation (RMP) coils to form the island divertor configuration [6]. The coils' magnetic spectrum is designed to generate RMP dominated by the $m/n$ = 3/1 and 4/1 components, where $m$ and $n$ are the poloidal and toroidal mode numbers. For $q_a \approx 3$, the coil system generates the 3/1 islands at plasma edge.

The formation process of the island divertor configuration is as follows: with externally imposed RMP to change the 3D magnetic field topology of the confined plasma, locked magnetic islands appear at the rational surfaces; when the magnetic field line near edge island separatrix is cut with a solid target plate (the HFS target), a structure of an open island region wrapping the residual islands is formed. In the experiment, to employ the HFS target as the primary limiter, the plasma is shifted 1 cm towards HFS, with a horizontal displacement of the plasma of -1 cm, making the minor radius $a$ = 0.215 m.

The 3D edge magnetic field topology is simulated in the vacuum assumption, without considering plasma responses. The 2D equilibrium magnetic field is computed by applying the equilibrium reconstruction code EFIT to J-TEXT. The background plasma parameters (e.g., the toroidal magnetic field, the plasma current, the major/minor radius, and the edge safety factor) in the simulation are consistent with the experiment. The 3D magnetic perturbation field by RMP coil is calculated by Biot-Savart's law. The total 3D magnetic field is the sum of the 2D equilibrium field and the 3D perturbation field. The field lines tracing (i.e., the step-by step integration of the field line equations) is used to numerically calculate the 3D magnetic topology.

The edge magnetic topology of the 3/1 island divertor configuration is calculated and shown in figure 1. The amplitude of the island divertor coil current $I_c$ is 4 kA and the phase of the magnetic field is set as negative to make the O-point of the edge island near the HFS target. The Poincaré map at the $\phi = 135°$ plane indicates that the field line near the edge island separatrix is intersected with the HFS target. And the 2D connection length $L_c$ distribution at the $\phi = 247.5°$ plane, which is the normal plane of the central view line of the CCD camera, shows the edge islands are divided into two regions: the open magnetic island region and the closed field line region. $L_c$ is defined as the length of one field line that starts from an initial point and ends with a striking point on the target or exceeding the tracing limit, set as 200 toroidal turns. The field lines are traced in both the forward and backward toroidal field directions. $L_c$ was calculated up to 500 m. Field lines with a total connection length $L_c \geq 500$ m are considered to be closed. The region inside the last closed flux surface (LCFS) and the residual 3/1 islands in SOL (scrape-off layer) are closed magnetic field regions, in which $L_c$ is effectively infinite ($L_c \geq 500$ m) since they are not intersected by the target plate. In the open region where $L_c$ is finite, field lines will reach the HFS target after a finite number of tracing turns.

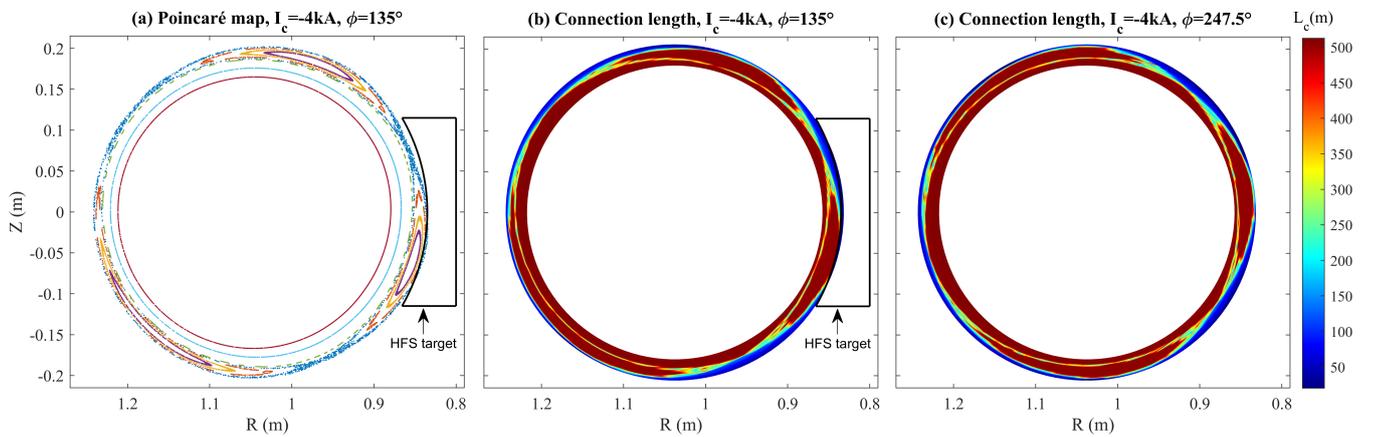

**Figure 1.** The edge magnetic topology of the 3/1 island divertor configuration on J-TEXT under $I_c$ = -4 kA. (a) The Poincaré map in the poloidal cross section of the port 6 where the HFS target is located ($\phi = 135°$). (b)-(c) 2D connection length distribution at the $\phi = 135°$ plane of the 3/1 island divertor configuration in two different poloidal cross sections, including (b) the cross section in the port 6 ($\phi = 135°$), (c) the normal plane of the central view line of the CCD camera ($\phi = 247.5°$).

*2.2. Formation of the island divertor configuration and Measurement of the island width on J-TEXT*

Several diagnostics for the detection of the formation of the island divertor configuration have been developed on J-TEXT. A limiter Langmuir probe array (LLPA) is embedded in the graphite tiles of the limiters to measure the floating potential ($V_f$) of the edge plasma [17]. A CCD diagnostic system is set up for visible light imaging [14]. The CCD camera is mounted at a tangential window of port 12, looking clockwise. The CCD tangential viewing scope can cover the whole cross-section of the J-TEXT tokamak. One cross-section corresponds to the normal plane of the central view line of the camera ($\phi = 247.5°$). A carbon III (CIII) filter is installed in front of the camera lens to obtain the spatial distribution of the CIII impurity radiation. Visible radiation from the plasma provides heat and light. The outline of the visible light can represent the edge magnetic island topology. Besides, the magnetic diagnostics of Mirnov probe arrays has been upgraded on the J-TEXT tokamak to measure the magnetohydrodynamic instabilities with higher spatial resolution and better amplitude-frequency characteristics. The Mirnov probe array contains one poloidal array with 48 probe modules and two toroidal arrays with 25 probe modules. Each probe module contains two probes which measure both the poloidal and the radial magnetic fields ($B_p$ and $B_r$) [15]. To measure the amplitude and phase of the non-axisymmetric radial magnetic field generated by the locked mode, 12 saddle loop sensors are developed on J-TEXT [16]. In the experiment, the saddle loop measures the radial magnetic field of locked mode. The magnetic diagnostics can give the amplitude of the $n$ = odd radial perturbation magnetic field ($b_r^{n=\text{odd}}$), dominated by the $n$ = 1 components. By combining these diagnostics, the formation of the island divertor configuration can be deduced.

The ECRH system has been designed and developed for J-TEXT tokamak to enhance the total heating power [13]. The system is composed of two gyrotrons and capable of providing power up to 1 MW. The launcher of ECRH is expected to inject the microwave power generated by the gyrotron to the desired position in the plasma. In the island divertor experiments, ECRH is deposited at $r = 0$ (on-axis) or 11 cm (off-axis), far from the $q = 3$ surface in the 3/1 island divertor configuration. The deposition location is changed by adjusting the poloidal injection angle.

For the experimental results described in this paper, the plasma parameters are as follows, unless otherwise noted: the major/minor radius ($R_0/a$) is 1.05 m/0.215 m, the toroidal magnetic field $B_t$ is 1.875 T, the plasma current $I_p$ is 120 kA, the edge safety factor $q_a$ is 3.1, and the core electron density ($n_e$) is approximately $3.1\times 10^{19}$ m$^{-3}$. An example of the formation of the 3/1 island divertor configuration is shown in figure 2 for shot #1088985. Before applying RMP, the plasma parameters remain nearly constant. The RMP coil current $I_c$ is initiated at $t = 0.2$ s, ramping up to the flattop of 5 kA (with a negative phase) at 0.23 s and is sustained until 0.51 s, followed by a decay to zero. The island divertor configuration is established at t = 0.226 s, as indicated by abrupt changes of the floating potential ($V_f$), and increases in the opened magnetic island width ($w$) and the radial magnetic perturbation field ($b_r^{n=odd}$). A more negative $V_f$ on the HFS target indicates that the field lines connected to the divertor target Langmuir probes penetrate deeper into the plasma core, and the heat load on the divertor target is higher. When the limiter configuration transitions to the island divertor configuration, the edge magnetic island is cut open by the HFS target, and the $V_f$ becomes more positive, suggesting that the field lines connected to the probes penetrate shallower into the plasma (near SOL), and hence the $L_c$ of the field lines connected to the target probes becomes smaller and the divertor heat-load is reduced. It should be emphasized that whether the edge island is opened by the target determines the formation of the island divertor configuration.

After the formation of the island divertor configuration, 200 kW of ECRH power is deposited at the plasma core ($r = 0$) from $t = 0.3$ s to 0.45 s. The main plasma parameters remain stable and the island divertor configuration can be sustained during the flattop of the RMP coil current.

When the coil current is decreased to 2.6 kA (at $t = 0.52$ s), $V_f$ changes to negative and the open island width $w$ becomes 0. The island divertor configuration transitions back to the limiter configuration. After that, as the coil current decreases, the radial magnetic field $b_r^{n=odd}$ reduces. When the coil current is decreased to 0 at $t = 0.526$ s, the $b_r^{n=odd}$ becomes 0. It should be noted that the measurements of the island size by the CCD and by the Mirnov probes are slightly different. The former one can measure the edge magnetic island width if and only if the island is cut open by the HFS target ($w > 0$, which determines the formation of the island divertor configuration), while the latter one can measure the radial perturbation field ($b_r^{n=odd} > 0$) generated by the edge island, even if the island is not opened by the divertor target ($w = 0$ in the limiter configuration).

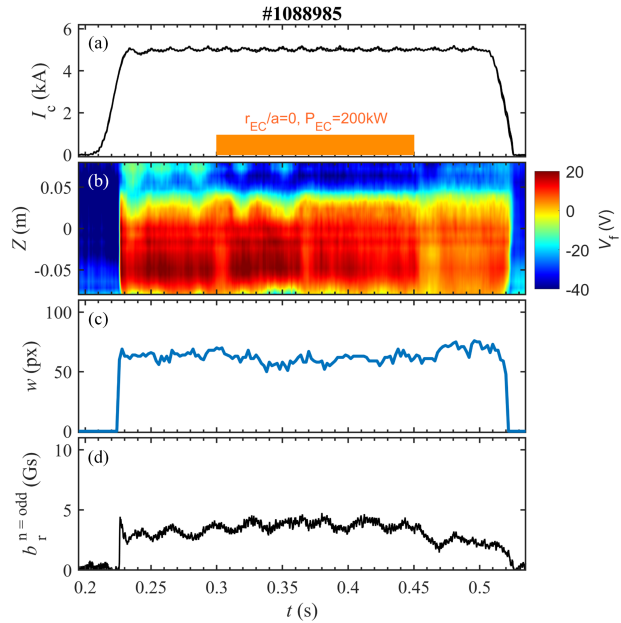

**Figure 2.** Overview of main plasma parameters in #1088985 shot on J-TEXT, including (a) the amplitude of the coil current $I_c$, and the ECRH power and deposition location, (b) the floating potential $V_f$ measured by the divertor Langmuir probes, (c) the edge island width $w$, obtained by the CCD image processing, (d) the amplitude of the $n$ = odd radial magnetic field ($b_r^{n=odd}$), measured by the Mirnov probes.

Figure 3 displays images captured by the CCD camera using CIII filter corresponding to figure 2. The outline of the visible light serves as a representation of the edge magnetic topology. Figure 3 (a) shows the CIII impurity radiation distribution of the limiter configuration. At $t$ = 0.2 s, the coil current is zero, and hence no magnetic island is formed.

Figure 3 (b) shows the CIII impurity radiation distribution of the island divertor configuration. The opened island region with a divertor-leg structure is observable at the HFS during the flattop of the coil current. When $I_c$ = -5 kA, the edge magnetic islands are excited by the RMP and cut open by the HFS divertor target. As a result, the motion of charged particles along the field lines in the open island region is interrupted by the target, breaking the closed particle trajectories. Notably, when ECRH is applied at $r$ = 0, the edge magnetic topology of the island divertor configuration remains stable.

The method for processing CCD images to calculate the open island width is illustrated in figures 3 (b1)-(b2). The width of the open edge magnetic island is determined by analyzing the gray-scale value distribution within the island region captured in the CCD image. The two bright lines of the open magnetic island are identified as two blue points, corresponding to the maximum gray-scale values along the orange line. By subtracting the pixel coordinates of these two points, the width of the open island is calculated. The unit of the width $w$ is in pixel. The evolution of $w$ shown in figure 2 is obtained by calculating the open island width in each frame of the CCD video. It should be emphasized that CCD can only detect the opened island. If the edge island is not cut open by the HFS target, the particle trajectories are closed, then the island cannot be detected by CCD.

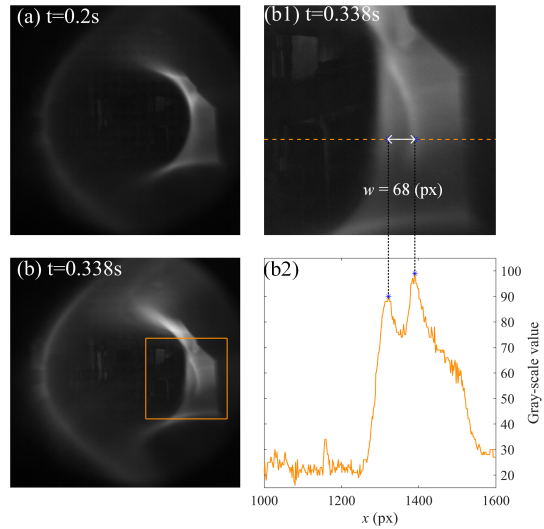

**Figure 3.** The CIII impurity radiation distribution of the limiter configuration and the island divertor configuration, corresponding to figure 2. (a) At $t$ = 0.2 s when the coil current $I_c$ = 0, the plasma is in the limiter configuration. (b) At $t$ = 0.338 s when the coil current $I_c$ = -5 kA, the plasma is in the island divertor configuration. (b1) The local image at the HFS shows the open island with a divertor-leg structure, corresponding to the rectangular box in figure 3(b). (b2) The gray-scale value distribution on the dashed line in figure 3(b1).

## 3. Investigation of the ECRH effect on the island divertor topology

*3.1. Observation of island healing induced by ECRH in the island divertor configuration*

Figure 4 displays a typical example of the island healing and the configuration transition induced by ECRH in shot #1088983. Following the formation of the island divertor configuration (not shown here for brevity), ECRH with a power of 600 kW is deposited in the plasma core starting at $t$ = 0.3 s. Upon the application of ECRH, the positive region of $V_f$ diminishes, becoming entirely negative at $t$ = 0.31 s. Concurrently, both $w$ and $b_r^{n=odd}$ reduce to zero at $t$ = 0.31s, indicating that ECRH induces the healing of the open island and the subsequent configuration transition. From $t$ = 0.31 s to $t$ = 0.45 s with ECRH, $V_f$ remains negative, $w$ remains zero, and $b_r^{n=odd}$ fluctuates around small values. The difference between $w$ and $b_r^{n=odd}$ ($w$ = 0, $b_r^{n=odd}$ > 0) suggests that the island still exists but is no longer opened by the target. When the island becomes closed, the width detected by the CCD is zero, while the magnetic probes still measure the perturbed fields generated by the island. Subsequently, upon turning off ECRH at $t$ = 0.45 s, $V_f$, $w$ and $b_r^{n=odd}$ all exhibit simultaneous and abrupt changes, signifying the recovery of the island divertor configuration.

When the deposition location remains unchanged, ECRH with lower power can have a negligible impact on the island divertor configuration. In shot #1088986, ECRH with a power of 400 kW is deposited at $t$ = 0.3 s. Under these conditions, $V_f$, $w$, and $b_r^{n=odd}$ remain stable. The island divertor images in this case are similar to Figure 3(b1) and hence will not be elaborated here.

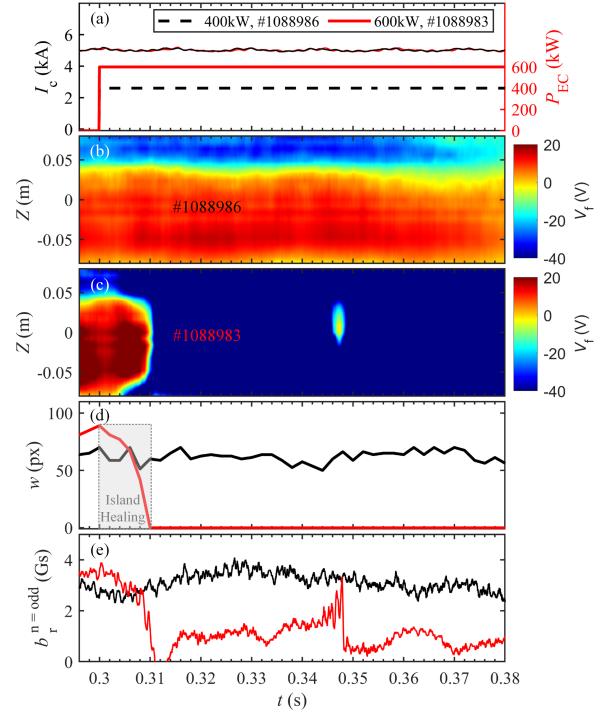

**Figure 4.** The J-TEXT #1088983 shot shows the phenomena of island healing and the configuration transition induced by ECRH with power of 600 kW, while the #1088986 shot with power of 400 kW does not show the healing phenomena. The main plasma parameters are: (a) the amplitude of the coil current $I_c$, and the ECRH power and deposition location, (b) the floating potential $V_f$ in shot #1088986, (c) the floating potential $V_f$ in shot #1088983, (c) the open island width $w$, obtained by CCD image processing, (d) the amplitude of the $n =$ odd radial magnetic field ($b_r^{n=odd}$), measured by the Mirnov probes.

Figure 5 illustrates the temporal evolution of the open island as observed by the CCD camera at various moments, corresponding to shot #1088983 shown in Figure 4. These images provide a more detailed visualization of the island healing process induced by ECRH. Between $t = 0.3$ s and $t = 0.31$ s, the CCD records a gradual reduction in the width of the open island. With an on-axis ECRH power of 600 kW, $\beta$ increases, and the open island tends to shrink. At $t = 0.31$ s, the island becomes closed, and the edge magnetic topology transitions to the limiter configuration. The heating duration for the configuration transition is approximately 0.01 s. Following the transition, the open island width remains at zero throughout the continued application of ECRH.

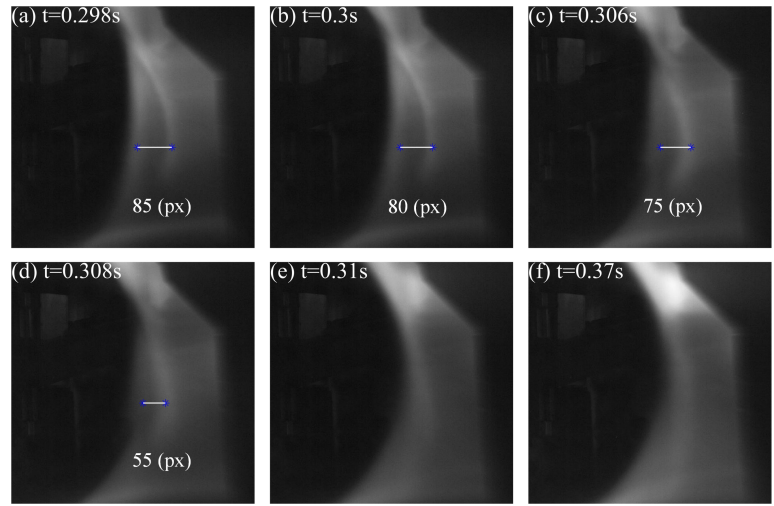

**Figure 5.** CCD images of island healing and configuration transition, corresponding to the #1088983 shot in figure 4. (a) At $t = 0.298$ s without ECRH, the plasma is in the island divertor configuration. (b)(c)(d) At $t = 0.3, 0.306$ and $0.308$ s, the open island tends to shrink with ECRH. (e) $t = 0.31$ s. The edge magnetic topology transitions to the limiter configuration. (f) $t = 0.37$ s. The limiter configuration remains.

Figure 6 presents another typical example of island healing and the critical phenomena associated with the configuration transition induced by ECRH. Following the formation of the island divertor configuration, ECRH with a power of 500 kW is deposited in the plasma core from $t = 0.3$ s to $t = 0.45$ s. Between $t = 0.3$ s and 0.32 s, the positive region of $V_f$ diminishes, while $w$ and $b_r^{n=odd}$ also decrease. At $t = 0.32$ s, $V_f$ becomes negative, and $w$ reduces to 0, the structure of the open island with a divertor-leg disappears. These facts show that the island divertor configuration transitions to the limiter configuration because of ECRH. The heating time for the transition is 0.02 s.

Interestingly, the critical phenomena in this shot differ from the previous results. An alternation between the island divertor and the limiter configuration is observed. At around $t = 0.34$ s, $V_f$ becomes positive, and both $w$ and $b_r^{n=odd}$ increase, indicating recovery of the island divertor configuration. After 0.01 s, it transitions back to the limiter configuration. This alternation occurs again at around $t = 0.36$ s, with the island divertor recovering momentarily before transitioning back to the limiter configuration within 0.01 s. At around $t = 0.445$ s, the island divertor configuration briefly reappears before transitioning back to the limiter configuration. After turning off ECRH, the 3/1 island divertor configuration fully recovers.

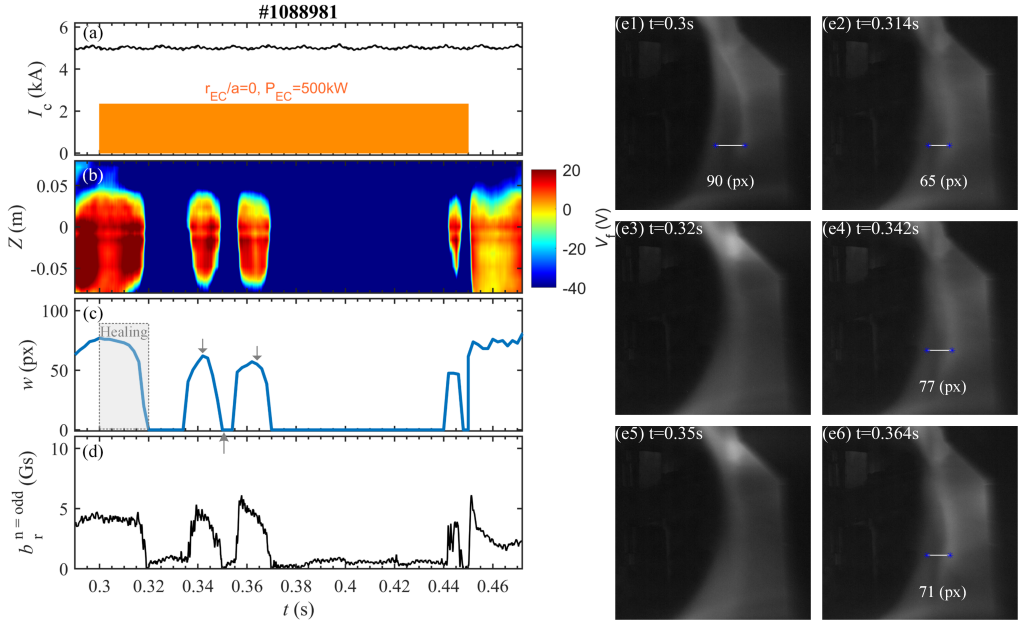

**Figure 6.** The J-TEXT #1088981 shot shows island healing and the critical phenomena of configuration transition induced by the ECRH with power of 500 kW. (a) the amplitude of the coil current $I_c$, and the ECRH power and deposition location, (b) the floating potential $V_f$ measured by the divertor Langmuir probes, (c) the open island width $w$, obtained by the CCD image processing, (d) the amplitude of the $n$ = odd radial magnetic field ($b_r^{n=odd}$), measured by the Mirnov probes. (e) CCD images of island healing and the critical phenomena of configuration transition at various moments: (e1) At $t = 0.3$ s, ECRH begins to be deposited in the island divertor configuration. (e2) At $t = 0.314$ s, the open island shrinks with ECRH. (e3) At $t = 0.32$ s, the edge magnetic topology transitions to the limiter configuration. (e4) At $t = 0.342$ s, the island divertor configuration recovers. (e) At $t = 0.35$ s, the open island disappears again. (f) At $t = 0.364$ s, the open island recovers.

If the width of the edge island is close to the critical width to be cut open by the HFS target and establish the island divertor configuration, even small fluctuations in the island size can trigger the configuration transition. This indicates that the topological structure of the open island becomes unstable. Consequently, it can be inferred that the magnetic island width decreases to near the critical width when ECRH with a power of 500 kW is deposited in the plasma core.

### 3.2. Lower power off-axis ECW deposition for the configuration transition

Figure 7 displays a typical example of the island healing induced by the off-axis ECRH in the island divertor configuration. After the formation of the island divertor configuration, ECRH is deposited at $r_{EC}$ = 11 cm ($r_{EC}/a$=0.5) with a power of 250 kW from $t = 0.37$ s to $t = 0.44$ s. At $t = 0.388$ s, $V_f$ changes to negative, while $w$ and $b_r^{n=odd}$ decrease to zero, indicating that the island divertor configuration transitions to the limiter configuration due to ECRH. Compared to the on-axis ECRH, off-axis ECRH with lower power is sufficient to induce the configuration transition.

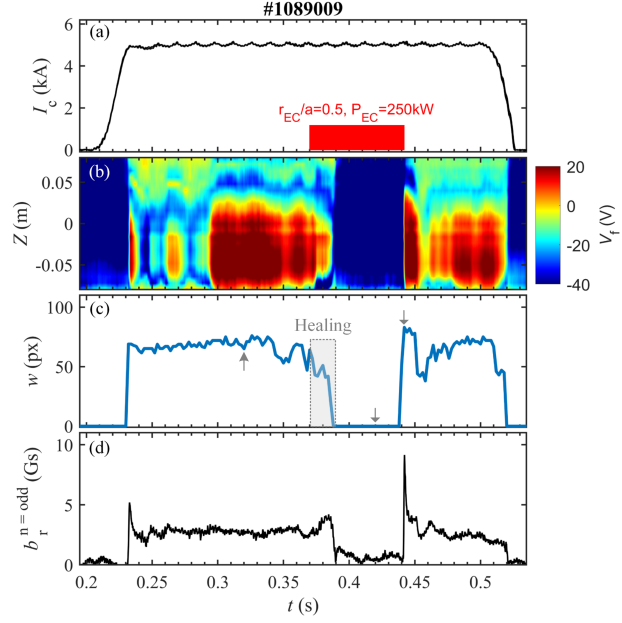

**Figure 7.** The J-TEXT #1089009 shot shows the phenomena of island healing and configuration transition induced by the off-axis ECRH heating with power of 250 kW. (a) the amplitude of the coil current $I_c$, and the ECRH power and deposition location, (b) the floating potential $V_f$ measured by the divertor Langmuir probes, (c) the edge island width $w$, obtained by the CCD image processing, (d) the amplitude of the $n$ = odd radial magnetic field ($b_r^{n=odd}$), measured by the Mirnov probes.

Figure 8 shows the phenomena of island healing induced by the off-axis ECRH. The width of the open magnetic island reduces from $t = 0.37$ s to $t = 0.388$ s. At $t = 0.388$ s, the structure of the open island with a divertor-leg disappears, marking the configuration transition. The heating time required for the configuration transition is 0.018 s. From $t = 0.388$ s to $t = 0.44$ s, the open island width remains at 0, sustaining the limiter configuration. Upon turning off ECRH, the island divertor configuration promptly recovers.

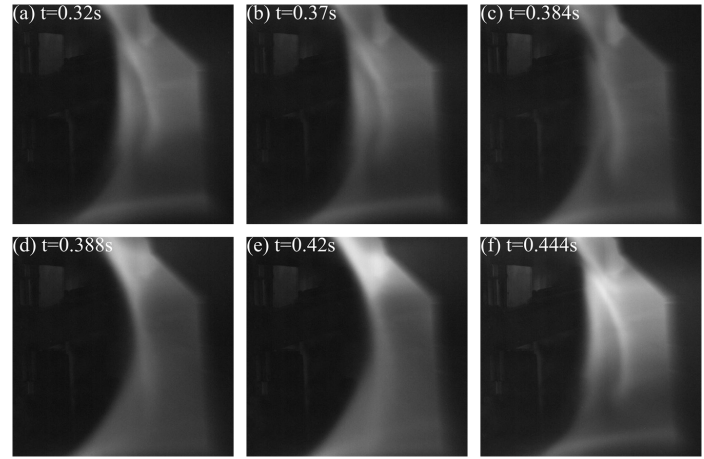

**Figure 8.** CCD images of island healing and configuration transition, corresponding to the #1089009 shot in figure 7. (a) At $t = 0.32$s, the plasma is in the island divertor configuration. (b) At $t = 0.37$s, ECRH begins to be deposited at $r = 11$cm. (c) At $t = 0.384$ s, the edge open island width reduces with the off-axis ECRH. (d) At $t = 0.388$ s, the open island disappears. (e) At $t = 0.42$ s, the plasma is still in the limiter configuration. (f) At $t = 0.444$ s, the island divertor configuration recovers after turning off ECRH.

The experiment results show that the off-axis deposition ($r_{EC}/a$=0.5) has a smaller power threshold for the island healing and the configuration transition compared to the on-axis deposition ($r_{EC}/a$=0). The results also indicate that the ECRH is better to be deposited at plasma core for a stable EC heated scenario in the island divertor configuration.

## 4. Discussion

The physical mechanism underlying the phenomena of island healing and the transition of the island divertor topology in J-TEXT induced by ECRH naturally warrants further investigation.

In stellarators, the healing of magnetic islands is attributed to plasma flows driven by neoclassical transport [11]. However, given the relatively smaller neoclassical transport in tokamaks, it is less probable for magnetic islands to heal via the same mechanism. Consequently, it becomes essential to develop a new model to explain the observed island healing in the island divertor configuration on J-TEXT.

Whilst there is uncertainty as to the mechanisms, there are several possibilities. One possible explanation is that ECRH, as a localized heat source, modifies the plasma temperature and alters the temperature difference between the two strike zones on the divertor target plate. This, in turn, affects the SOL current flowing along the open magnetic island, ultimately leading to changes in the island topology.

Figure 9 illustrates the two strike points on the HFS target resulting from one of the finite-$L_c$ field lines, along with the path of the SOL current flowing through the helical flux tube. The field lines originating from the open island region reach the HFS target after a finite number of tracing turns.

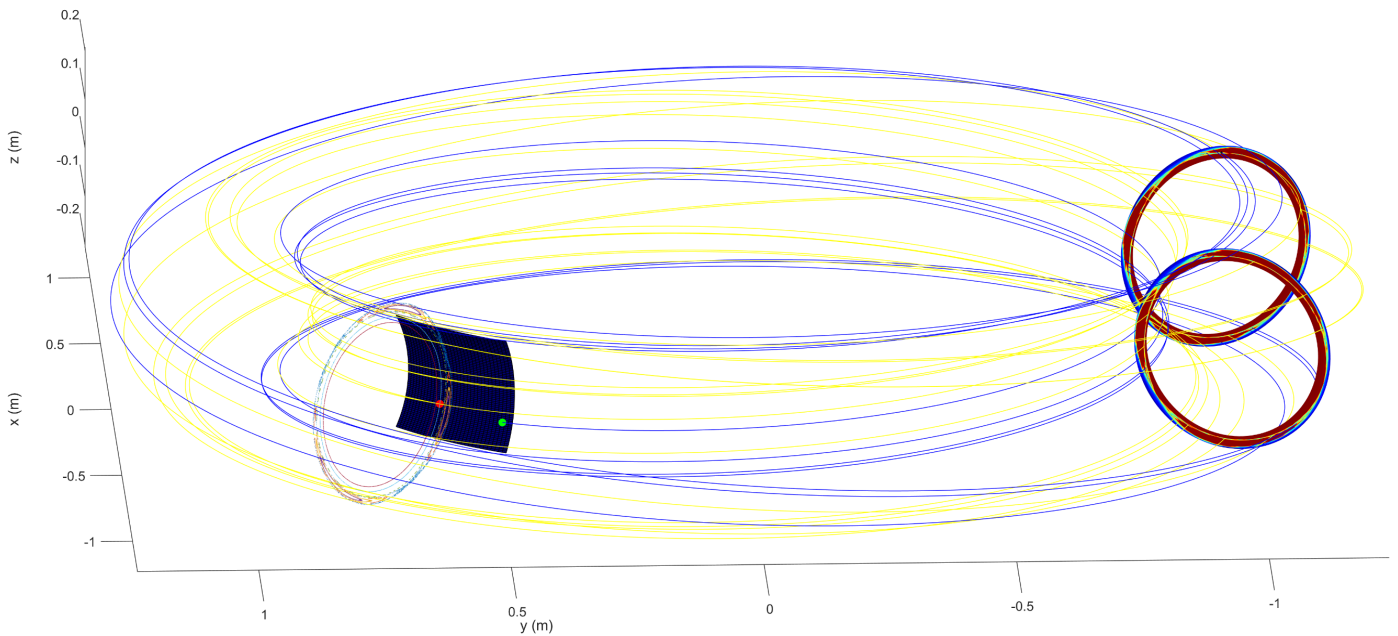

**Figure 9.** Field line tracing in the 3D Cartesian coordinate system in the 3/1 island divertor configuration under $I_c$ = -4 kA, combined with the Poincaré map in the poloidal cross section of the port 6 where the HFS target located ($\phi$ = 135°), and 2D $L_c$ distribution at different toroidal angles ($\phi$ = 247.5° and 292°).

The trajectory of the field line is obtained by integrating the magnetic field line equation, using 4th Runge-Kutta method. Tracing is performed in both clockwise and counterclockwise directions, starting from an initial point in the open island region. After a finite number of toroidal turns, the field lines terminate at two striking points on the target. The field line traced in blue follows the clockwise direction and ends at the green striking point. Conversely, the field line traced in yellow follows the counterclockwise direction and terminates at the red striking point, which is located higher than the green point.

In the island divertor configuration, a temperature difference exists between the two striking zones on the divertor target plate. As illustrated in figure 10, particles carrying heat mainly deposit on the upper strike line, leading to a higher temperature on the upper strike line compared to the lower one. This temperature difference between the two ends of the SOL can drive currents in the SOL of tokamak plasmas through thermoelectric effects, as demonstrated in previous research [18].

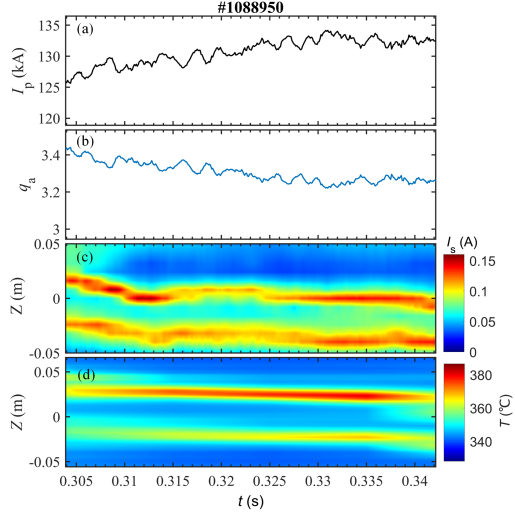

**Figure 10.** Asymmetry of the energy deposition patterns on the HFS target plate in the island divertor configuration. $I_c$ = 5 kA. (a) the plasma current $I_p$, (b) the edge safety factor $q_a$, (c) the ion saturation current $I_s$ measured by the divertor Langmuir probes, (d) the temperature on the HFS divertor target plate, measured by the infrared camera.

If the main striking zone on the divertor target plate is under temperature $T_1$, and the second striking zone is under temperature $T_0$ ($T_1 > T_0$), then a thermoelectromotive force $E_T$ can be driven by the temperature difference ($T_1 - T_0$). The equation is $E_T = \alpha(T_1 - T_0)$.

The Spitzer resistivity of hydrogen plasma is given as $\sigma_\| = 5.2 \times 10^{-4}/(T^{3/2}$ [eV]) (Ω-m). The flux tube with a finite connection length $L_c$ between the two SOL ends on the HFS divertor target has a finite resistivity $R = \sigma_\| \cdot L_c$. Therefore, the thermoelectric current within the SOL flux tube can be given as $I_T = E_T/R$.

$$I_T = \alpha(T_1 - T_0)/[L_c \cdot 5.2 \times 10^{-4}/(T^{3/2})] = \frac{\alpha(T_1 - T_0) \cdot T^{3/2}}{5.2 \times 10^{-4} \cdot L_c}$$

ECRH can lead to the increase of the SOL thermoelectric current, based on the following illation. With the deposition of ECRH heating, the plasma temperature is increased, and hence the Spitzer resistivity of plasma $\sigma_\|$ is decreased and the resistivity $R$ of the finite-$L_c$ helical flux tube in the SOL plasma becomes smaller. And due to the increased temperature after applying ECRH, assume $T_1' \approx k_1 T_1$ and $T_0' \approx k_0 T_0$, then the temperature difference ($T_1' - T_0'$) between the two SOL ends on the HFS divertor target plate becomes higher, $T_1' - T_0' \approx k_1 T_1 - k_0 T_0$. Thus, the thermoelectromotive force $E_T$ is increased. Therefore, the thermoelectric current within the SOL flux tube $I_T$ is increased by ECRH.

$$\Delta I_T = \frac{\alpha(T_1' - T_0') \cdot T'^{3/2} - \alpha(T_1 - T_0) \cdot T^{3/2}}{5.2 \times 10^{-4} \cdot L_c}$$

The magnetic perturbation field generated by the increasing SOL thermoelectric current and the change of magnetic topological structure of plasma boundary can be calculated. The simulation results show that the magnetic perturbation generated by the SOL helical currents can induce changes in the edge magnetic topology of the island divertor configuration. The effect of ECRH on the SOL thermoelectric current can partly explain the edge island width reduction and even the configuration transition, based on the following numerical calculations.

Firstly, the paths of the SOL filament currents are determined through field line tracing. One example of the path of the finite-$L_c$ flux tube in SOL plasma, carrying the thermoelectric current $I_{SOL1}$, is shown in figure 9. The field line originating from the open region of the edge magnetic island ends with two striking points on the HFS divertor target plate. The current flows from the primary striking zone, which is at a higher temperature ($T_1$), which is at a lower temperature ($T_0$).

Next, the magnetic perturbation field generated by the SOL thermoelectric current is calculated. Phases of the $m = 3$, $n = 1$ magnetic perturbation generated by the four SOL filament currents and by the island divertor coil current are shown in figure 11 (c). The four SOL filament thermoelectric currents all flow within flux tubes characterized by short connection lengths ($L_c$). Furthermore, figure 11 (a) presents the 2D $L_c$ distribution, while figure 11 (b) shows the 1D $L_c$ distribution along the dashed line in (a). The Z-coordinates of the four SOL currents, when they pass through the dashed line at $R = 1.05$ m, $\phi = 292°$, are marked as four points in the 1D-$L_c$ plot.

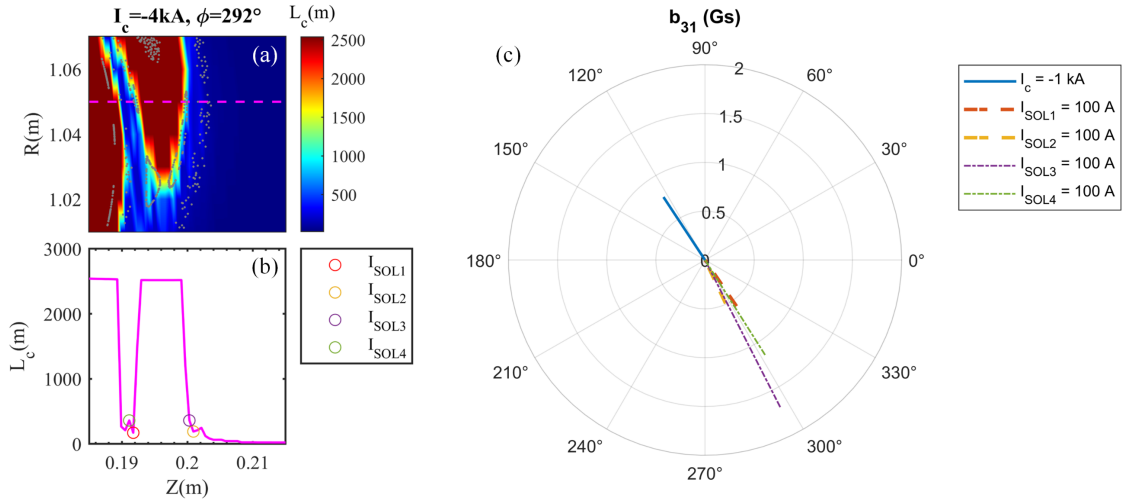

**Figure 11.** (a) 2D distribution of the field line connection length $L_c$. (b)1D distribution of $L_c$ along the dashed line in (a). The points of the four SOL filament currents are labeled in the 1D-$L_c$ plot. The four SOL filament thermoelectric currents flow in the flux tubes with short connection length. (b)Phases of the *m*=3, *n*=1 magnetic perturbation generated by the four SOL currents (100 A) and by the coil current (-1 kA).

The phases of the $m = 3$, $n = 1$ component of the magnetic perturbation field generated by the filament currents within the short-$L_c$ flux tubes in the SOL plasma are nearly opposite to the phase of the $m/n = 3/1$ perturbation generated by the island divertor coil current. Therefore, the SOL filament currents can counteract the effect of the IDC current and reduce the size of the $m/n = 3/1$ edge open island. For example, $I_{SOL1} = 100$ A can counteract $I_c = -0.75$ kA, as illustrated in figure 11 (c). According to Biot-Savart's law, when the SOL current is located near the edge island, even relatively small currents can have a significant impact on the edge island topology.

The spectral components of the magnetic perturbation field produced by the four SOL currents are calculated separately. An example of these calculations is presented in figure 12. The calculation method is based on spectrum analysis [19].

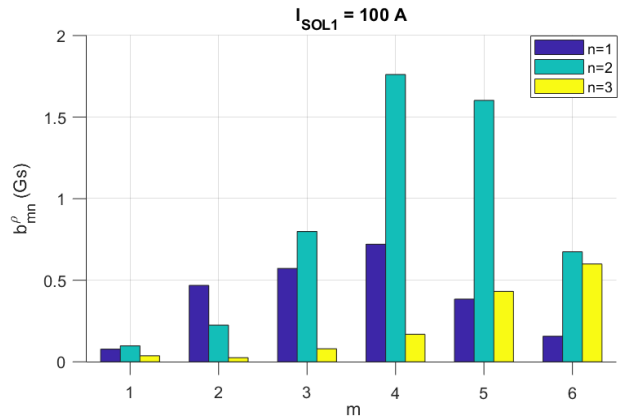

**Figure 12.** Spectral components of the radial magnetic field generated by the SOL filament current $I_{SOL1}$.

Healing of the opened edge magnetic island due to the thermoelectric currents in the SOL is calculated through numerical modelling. The SOL filament current $I_{SOL1}$ is selected to do the calculations of the change of magnetic topological structure of plasma boundary.

The simulation results in figure 13 demonstrate that increasing the SOL filament current has an effect equivalent to decreasing the amplitude of the island divertor coil current. The $L_c$ distribution with an SOL current of 0.1 kA and a coil current of -5 kA closely resembles the $L_c$ distribution with a coil current of -4 kA. Both increasing the SOL current and reducing the amplitude of the coil current lead to a shrinkage of the 3/1 magnetic island, resulting in a smaller open island region.

If the area of the open island region reduces to zero—meaning the edge island no longer intersects the HFS target—then the island divertor configuration transitions to the limiter configuration. Figure 13 (c1) shows the critical point at which this configuration transition occurs. The simulation successfully reproduces the phenomena that the open island is healed by ECRH. It can therefore be inferred that the effect of ECRH on the SOL thermoelectric currents within the finite-$L_c$ flux tubes can serve as an explanation for some of the experimental observations.

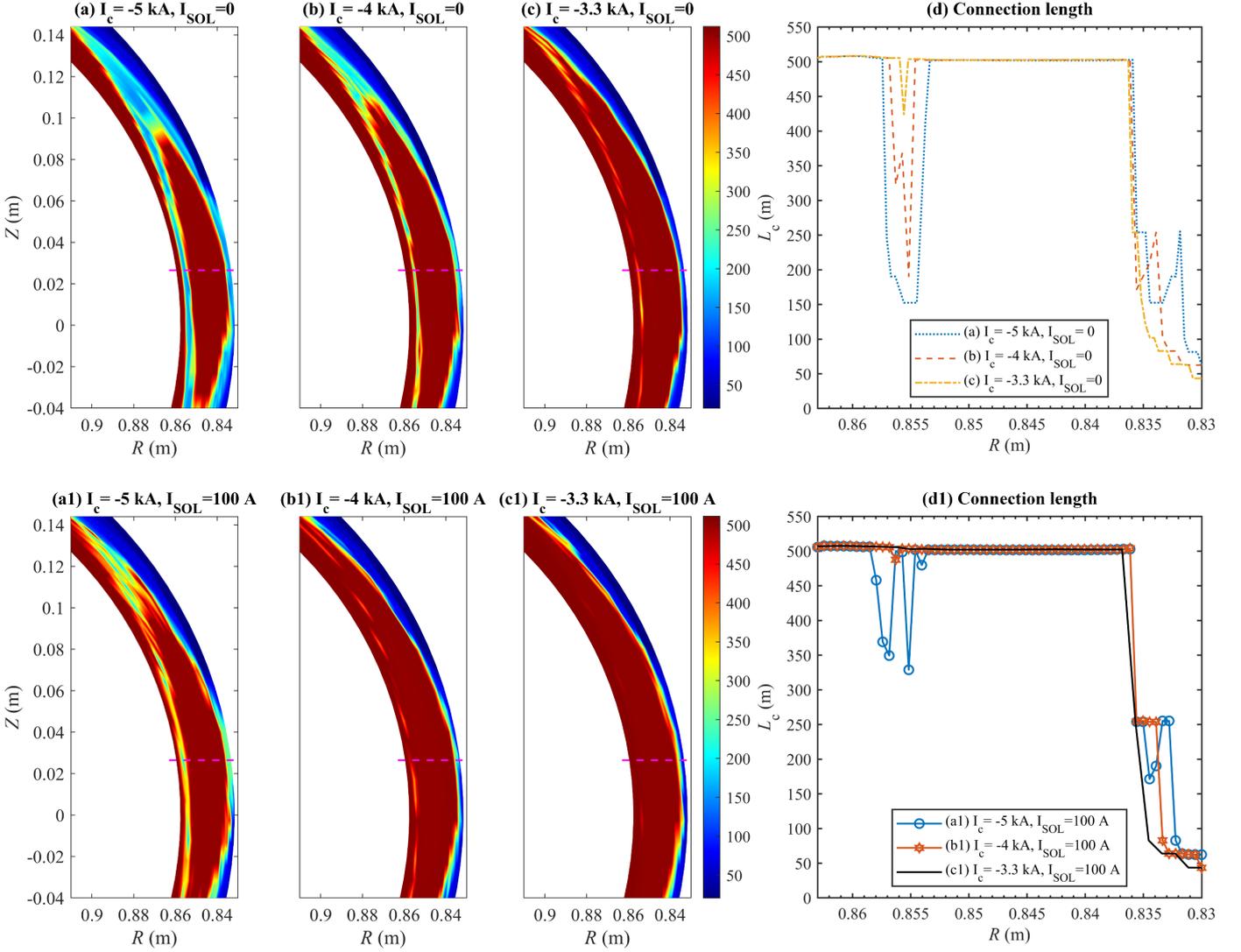

**Figure 13.** 2D distribution of the connection length $L_c$ at the edge 3/1 island, and 1D $L_c$ distribution along the dashed line in the 2D-$L_c$ plot, with different $I_c$ and $I_{SOL}$. The upper row is without the SOL current. (a) $I_c$ = -5 kA, $I_{SOL}$ = 0, (b) $I_c$ = -4 kA, $I_{SOL}$ = 0, (c) $I_c$ = -3.3 kA, $I_{SOL}$ = 0. The lower row is with the SOL current. (a1) $I_c$ = -5 kA, $I_{SOL}$ = 100 A, (b1) $I_c$ = -4 kA, $I_{SOL}$ = 100 A, (c1) $I_c$ = -3.3 kA, $I_{SOL}$ = 100 A.

## 5. Conclusion

In this report, the phenomena of island healing and the configuration transition is discussed. In the island divertor configuration on the J-TEXT tokamak, the size of the edge opened magnetic island decreases substantially, when ECRH with power of 500~600 kW is deposited at the plasma core, or ECRH with 250 kW is deposited at $r_{EC}$ = 0.5 $a$. In both scenarios, the island divertor configuration is no longer effective and transitions to the limiter configuration due to the reduced island size. Notably, on-axis deposition exhibits a higher power threshold for inducing island healing and the configuration transition. In the J-TEXT island divertor experiment, the plasma volume was estimated to be $V = 2\pi R_0 \cdot \pi a^2 \approx 1$ m$^3$. For a core deposition of 500 kW ECRH power, the corresponding power density is 500 kW/m$^3$. Scaling this to the W7-X island divertor, with a plasma volume of ~30 m$^3$, would imply a power deposition of approximately 500·30 kW = 15 MW. The power threshold of the off-axis ECRH would be even lower. These suggest that the deposition location and the heating power of ECRH should be controlled to achieve robust operation of the island divertor configuration.

Numerical simulations of the edge magnetic topology in the island divertor configuration were also conducted, successfully reproducing the observed island healing phenomena by considering the SOL thermoelectric currents increased by ECRH.

Although the initial experimental results are promising, further systematic studies are still ongoing. Specifically, the performance of the 4/1 island divertor configuration in the high-power plasma operation needs to be analyzed. Moreover, the role of increased helical SOL currents driven by ECRH, which serves as one potential explanation for the observed phenomena, requires further experimental validation. Future efforts will aim to explore alternative physical mechanisms and conduct comprehensive analyses to enhance the understanding and optimization of the island divertor configuration under high-power plasma operation.

## Acknowledgments

The authors extend their gratitude to journal reviewers and the J-TEXT team for their invaluable assistance.

This work is supported by the National Natural Science Foundation of China (No. 12305243 and 12375217) and Hubei Provincial Natural Science Foundation of China (No. 2022CFA072).